\documentclass{article}

\usepackage{arxiv}
\usepackage{algorithm}
\usepackage{algorithmic}
\usepackage{microtype}
\usepackage{graphicx}
\usepackage{subfigure}
\usepackage{booktabs} 
\usepackage{biblatex}
\usepackage[utf8]{inputenc}
\usepackage{hyperref}

\usepackage{amsmath}
\usepackage{amssymb}
\usepackage{mathtools}
\usepackage{amsthm}

\usepackage[capitalize,noabbrev]{cleveref}

\theoremstyle{plain}
\newtheorem{theorem}{Theorem}[section]

\theoremstyle{definition}

\newtheorem{assumption}[theorem]{Assumption}
\theoremstyle{remark}
\newtheorem{remark}[theorem]{Remark}

\usepackage[textsize=tiny]{todonotes}

\usepackage{xcolor}
\newcommand {\V}[1] {{\mbox{\boldmath $#1$}}}
\newcommand{\R}{\ensuremath{\mathbb{R}}}
\newcommand{\GS}{G^{S}}
\newcommand{\GA}{G^{A}}
\newcommand{\A}{\V{A}}
\usepackage{tabularx}
\DeclareMathOperator*{\argmin}{arg\,min}
\DeclareMathOperator*{\Tr}{Tr}

\title{Weakly Supervised Indoor Localization via Manifold 
Matching}

\author{ \href{https://orcid.org/0000-0002-8854-8497}{\includegraphics[scale=0.06]{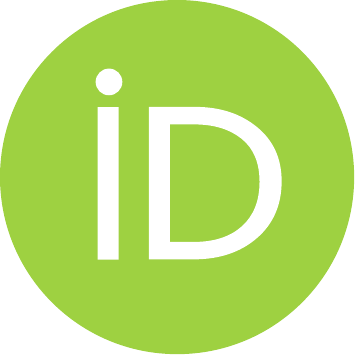} \hspace{1mm} Erez Peterfreund } \\
	Department of Computer Science\\
	The Hebrew University of Jerusalem\\
	Jerusalem 9190401, Israel\\
	\texttt{erezpeter@cs.huji.ac.il} \\
	\AND
	 \hspace{1mm} Ioannis G. Kevrekidis \\
	Department of Chemical and Biomolecular Engineering\\
	Johns Hopkins University\\ Baltimore, MD 21218\\
	\And
	 \href{https://orcid.org/0000-0001-7637-7128
}{\includegraphics[scale=0.06]{orcid.pdf} \hspace{1mm}Ariel Jaffe} \\
	Department of Statistics and Data Science\\
	The Hebrew University of Jerusalem,\\ Jerusalem 9190401, Israel\\
	\texttt{ariel.jaffe@mail.huji.ac.il} 
}

\date{}

\renewcommand{\undertitle}{}

\renewcommand{\shorttitle}{Weakly Supervised Indoor Localization via Manifold 
Matching}

\hypersetup{
pdftitle={A template for the arxiv style},
pdfsubject={q-bio.NC, q-bio.QM},
pdfauthor={David S.~Hippocampus, Elias D.~Striatum},
pdfkeywords={First keyword, Second keyword, More},
}

\begin{document}
\maketitle

\begin{abstract}
    Inferring the location of a mobile device in an indoor setting is an open problem of utmost significance. 
    A leading approach that does not require the deployment of expensive infrastructure is \textit{fingerprinting}, where a classifier is trained to predict the location of a device based on its captured signal. The main caveat of this approach is that acquiring a sufficiently large and accurate training set may be prohibitively expensive.
    Here, we propose a weakly supervised method that only requires the location of a small number of devices.
    The localization is done by matching a low-dimensional spectral representation of the signals to a given sketch of the indoor environment. 
    We test our approach on simulated and real data and show that it yields an accuracy of a few meters, which is on par with fully supervised approaches. 
    The simplicity of our method and its accuracy with minimal supervision makes it ideal for implementation in indoor localization systems.
\end{abstract}

\section{Introduction}

The development of algorithms for indoor positioning based on Wi-Fi or cellular signals has gained considerable interest in recent years due to their applicability to location-based services. These services include, for example, network management, security,  healthcare, and emergency navigation as well as many commercial applications \cite{zafari2019survey,khalajmehrabadi2017modern}. A typical setting includes a small number of receivers that capture signals transmitted by multiple mobile devices whose location is unknown. The task is to infer the position of these mobile devices solely based on the captured signals.  

\begin{figure}[tb]
    \centering
    \includegraphics[width=.48\columnwidth]{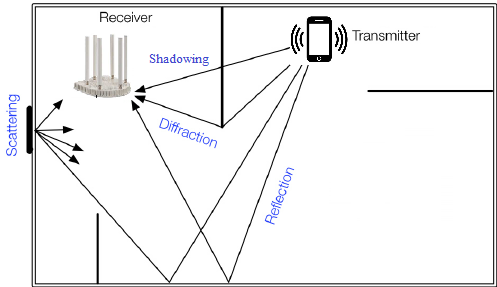}
    \includegraphics[width=.48\columnwidth]{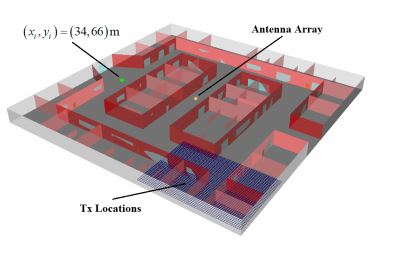}
    \caption{Left: An illustration of a signal's propagation paths from a cellphone to the receiver.  Right: The office building we use in our experiment in \cref{sec:experiment1}.
    }    
    \label{fig:indoor_positioning_setting}
\end{figure}

In an outdoor setting, the problem is readily solved by using triangulation based on GPS signals. This approach is applicable since the transmitter (e.g., satellite ) has a direct Line-Of-Sight connection with the users (e.g., cellphones). In an indoor setting, however, the transmitted signal is usually reflected, scattered, shadowed, or diverted from the Line-Of-Sight \cite{witrisal2012performance}. The result of these diversions is the ``multipath'' phenomenon where the transmitted signal propagates through different paths to the receiver, see illustration in \cref{fig:indoor_positioning_setting}. The problem of indoor localization thus requires a different approach. A hardware-oriented method to enable positioning involves distributing a large number of receivers throughout the indoor area. This will mitigate the ``multipath'' phenomenon and enable positioning through triangulation by the closest receivers. The caveat of this method is the high cost and complexity of deploying such a large network of receivers \cite{liu2014survey,ni2003landmarc}.

An alternative, data-oriented approach that does not require expensive infrastructure is called \textit{fingerprinting}. Here, the task of indoor positioning is cast as a supervised learning problem. First, a training set is obtained by capturing signals through the receivers while positioning a transmitter at multiple \textit{known} locations. From each transmitted signal, we extract a \textit{fingerprint}, a set of features of the signal that characterizes the transmitter's location. Common features include, among others, the signal's strength, \cite{barsocchi2009novel}  the channel information \cite{yang2013rssi} and multi-path time and direction of arrival \cite{wang2016tdoa,kupershtein2012single,jaffe2014single} . This dataset is used to train a classifier that predicts, for new  captured signals, the location from which they were transmitted \cite{chen2017confi}. 

One major problem of the fingerprinting approach for applications requiring high accuracy is that recording
and maintaining a suitable labeled data set may be prohibitively expensive. In contrast, collecting even vast amounts of unlabeled data may be done simply by recording the Wi-Fi signals of various devices moving through the venue. This fact motivates a ``weakly supervised fingerprinting'' problem, in which, except for a small number of points, the given data includes only the received signals without their transmitted locations. The objective is to estimate the location from which the signals were transmitted.

\begin{figure}[t]
    \centering
    \includegraphics[width=.48\columnwidth]{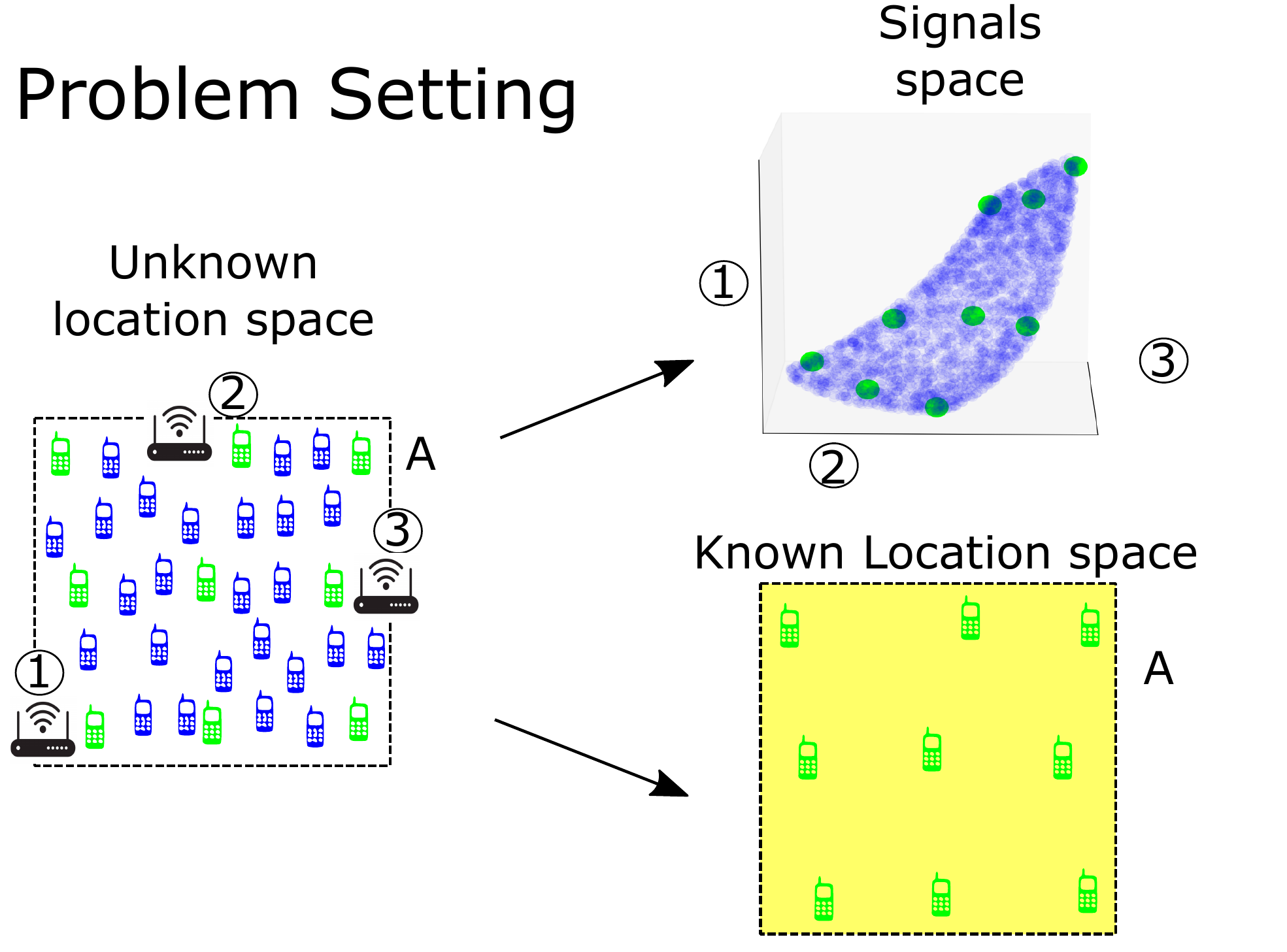} 
    \vline
    \includegraphics[width=.48\columnwidth]{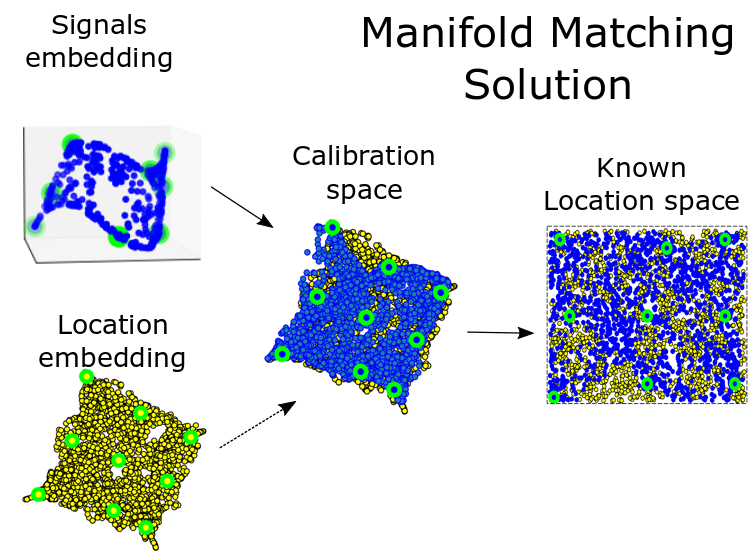} 
    \caption{ Left: Problem illustration- multiple devices are transmitting signals from different locations in an indoor environment, which are captured by $3$ Wifi receivers. Given the entire set of captured signals, the location of the green devices and the shape of the environment - the task is to estimate the locations of the blue devices. Right: A spectral  embedding is extracted from both the signals and points sampled from the known environment shape. Then, the location of the green devices is used to calibrate the signal's embedding by matching it with the location embedding. The calibrated embedding is then used for localization.}
    \label{fig:framework}
\end{figure}

This work aims to tackle this semi-supervised problem while assuming that the shape of the indoor environment is provided. Our approach is based on matching a spectral representation of the captured signals to a spectral representation of the given indoor environment. 
To that end, we use a small number of signals whose location is known as calibration points that enable accurate matching between the two representations. The problem, along with our approach for its solution, is illustrated in \cref{fig:framework}.

This paper proceeds as follows: In \cref{sec:related work} we provide a line of relevant work and describe how our setting relates to other problems of semi-supervised learning. In \cref{sec:problem settings} we formally define our framework and our model assumptions. The steps for our manifold-matching approach for indoor localization are outlined in \cref{sec:solution}. Experiments on simulated and real data are presented in  \cref{sec:experiments}. The paper concludes with  \cref{sec:discussion}, where we discuss potential improvements and future work.

\section{Related work}

\label{sec:related work}

The weakly supervised Wifi localization problem that we tackle in this paper is, at its core, a semi-supervised problem. The task is to estimate the locations (labels) of a large set of unlabeled signals given a small subset of labeled ones. The classic semi-supervised setting is addressed in \cite{belkin2004semi,belkin2004regularization,belkin2006manifold} where the goal is to compute a low-dimensional representation for the given data that is smooth with respect to a computed graph but also similar, for a subset of points, to their provided labels.
Similar semi-supervised methods were used, among others, for image and text classification  \cite{zhou2004learning, fergus2009semi}, see extensive review in \cite{zhou2014semi}. 

The main difference between our problem and the classic semi-supervised settings is that we are given, as side information, the shape of the area from which the points were transmitted. 
As we show, under appropriate assumptions, it is possible to leverage the geometric structure of the area to compute a correspondence between the signals and the location from which they were transmitted.

Challenges that include matching signals to a given shape exist in many other domains. In  \cite{villoutreix2017synthesizing} the goal is to perform matrix completion by computing a map between two trajectories.
Another example is the explainability of latent variables \cite{koelle2018manifold}. Here, the 
goal is to unveil  connections between the graph's eigenvectors and user-defined functions that originate from domain knowledge.  
In computer vision, a classic challenge is to computing a correspondence between $2d$ and $3d$ surfaces \cite{rodola2017partial,litany2017fully}.  

For cases where additional local information is available (in the form of systematic small perturbations), anisotropic diffusion maps can be used to  map the signal space to the floor plan geometry \cite{singer2008non}. This matching can be accomplished via an orthogonal transformation between the two embeddings, needing very few known registration points. A neural network architecture implementing a similar procedure has been published in \cite{peterfreund2020local}. Here, the neural network find an embedding to the signal space that matches, up to orthogonal transformation, to the floor plan map. 

For the application of indoor localization,  a semi-supervised approach was derived in \cite{moscovich2017minimax}, where a signal's location was estimated based on its geodesic distance to the closest signal whose location is known. In a recent paper, \cite{ghazvinian2021modality} derived an approach for indoor localization that is based on the given shape of the indoor environment. Their algorithm computes a map between the signals and their locations by optimal transport. 
This approach, however, requires a partial label in the form of a room or zone of the given signals. In contrast to \cite{ghazvinian2021modality} , our approach computes a transformation between a graph-based representation of the signals, and a graph-based representation of the reference floor plan.
We show that a small number of signals with known locations suffice for computing this transformation and obtaining localization accuracy of a few meters.

\section{Problem  setting}
\label{sec:problem settings}
We consider an array of $p$ receivers and $M$ transmission devices located in an area denoted by $A\subset \mathbb{R}^2$. For example, $A$ may be a single floor area in an office building, a mall, or an airport. 
Each device transmits a set of $K$ signals from approximately the same location, which are captured by the $p$ receivers. We denote the location of the $i-$th device by $\V{x}_i\in A$ and the set of its transmitted signals  by $\V{S}^{i} = [\V{s}_1^i,\ldots,\V{s}_K^i] \in \mathbb{C}^{K \times p}$. We assume that the locations of a small number of devices $\V{x}_1,\ldots,\V{x}_N$ are known, where $N \ll M$. These can be, for example, the location of stationary transmitters that repeatedly communicate with the receivers.
The locations $\V{x}_{N+1},\ldots,\V{x}_M$ of the rest of the $M-N$ devices are unknown, and our goal is to provide an accurate estimate of them. \cref{fig:framework} illustrates our setting.

Clearly, this task is not possible without  additional assumptions. Here, we make the following  assumptions on the input signals.   
 \begin{assumption} \label{assum:uniform}
  \textit{Uniform sampling}- we denote by $\V{X}$ a random variable with a uniform distribution over $A$. The locations $\{\V{x}_i\}$ are independent realizations of $\V{X}$. 
 \end{assumption}
 \begin{assumption} \label{assum:sufficient}
  \textit{Sufficient Sampling} - $M$ is sufficiently large to cover the entire area $A$ with high resolution. 
 \end{assumption}
 \begin{assumption} \label{assum:smoothness}
    \textit{Smoothness} - 
the signal $\V{s}_j^i$ is a  bilipschitz function of the location $\V{x_i}$ and a small set of parameters $\theta_i$, that are independent of the locations. Example for such parameters include transmittor type, orientation etc. Thus, the signals forms a low dimensional manifold embedded in a $K\cdot p$-dimensional space.
 \end{assumption}
The above assumptions are standard in the literature of indoor positioning \cite{jaffe2014single,ghazvinian2021modality}. In \cref{sec:discussion}, we discuss the common case where a physical barrier in the indoor enviornment causes the signal to be discontinuous. 
In section \ref{sec:solution} we leverage our knowledge of $A$ as well as the location of the $N$ transmitters to infer the location of the other devices.

\begin{algorithm}[tb]
\caption{Indoor localization by Manifold Matching}\label{alg:localization}
\begin{algorithmic}[1]
\STATE {\bfseries Input:}\begin{tabular}[t]{ll}
            $A\in \mathbb{R}^2$ &Floor plan \\
            $\{\V{S}^1,\ldots, \V{S}^M\}\subset \mathbb{S}$ & Transmitted signals \\
         $\{\V{x}_1,\ldots,\V{x}_N\}\subset A$ &$N$ Known signals \\
            & locations ($N\ll M$) \\
            $K:\mathbb{S}\times \mathbb{S} \rightarrow \mathbb{R}_{+}$ & Signals similarity function \\
            $d,l$ & Embedding dimensions.
\end{tabular}
\STATE {\bfseries Output:}\begin{tabular}[t]{ll}
            $\hat {\V{x}}_{N+1},\ldots,\hat {\V{x}}_{M}\in \mathbb{R}^2$ &  Estimated location \\
            & for signals
\end{tabular}

\STATE \underline{Generate a representation for the signals:}
\begin{tabular}[t]{l}
- Build a graph on the signals $\GS$ by applying $K$ on $\{\V{S}^i\}$\\
- Compute the normalized Laplacian $\V{L}^{S}$, as in Eq. \eqref{eq:rw_laplacian}\\
- Generate an embedding $\V{\phi}_S\in \mathbb{R}^{M\times d}$   as in Eq. \eqref{eq:embedding}
\end{tabular}
\STATE \underline{Generate a spectral representation for $A$:}
\begin{tabular}[t]{l}
- Sample $\V{y}_1,\ldots,\V{y}_{T}$ uniformly from $A$ \\
- Build a graph on the sampled locations $\GA$ \\
\,\, using a normalized Gaussian kernel\\
- Compute the normalized Laplacian $\V{L}^{A}$\\
- Generate an embedding $\V{\phi}_A\in \mathbb{R}^{T\times l}$,  as in Eq. \eqref{eq:embedding}
\end{tabular}
\STATE Generate a calibrated signal embedding as  shown in Eq. \eqref{eq:calibrated_embedding_form}, $\V{\psi}_1,\ldots,\V{\psi}_M$
\STATE Estimate the location of the transmitted signals in $A$ via $1$nn as shown in Eq. \eqref{eq:estimate_loc}.
\end{algorithmic}
\end{algorithm}

\section{Indoor localization via manifold matching}
\label{sec:solution}
In this section we derive the steps of our localization approach in detail. 

\paragraph*{A tale of two graphs.} 
Constructing a neighborhood graph is a common approach in manifold learning. Here, we compute two such graphs. The nodes of the \textit{signal graph}, denoted by $G^{S}=(V^{S}, E^{S}, W^{S})$, correspond to the $M$ transmitting devices. The weight between a pair of nodes $v_i^S, v_j^S\in V^{S}$ is calculated based on a predefined local kernel function $K(\V{S}^i,\V{S}^j)$. The choice of $K(\cdot,\cdot)$ depends on the characteristics of the captured signals.  
Importantly, the smoothness assumption implies that pairs of nodes connected by an edge with a large weight correspond to devices  located at close locations within $A$. 

The second graph, denoted $G^{A}=(V^{A},E^{A}, W^{A})$ is the \textit{area graph}, which is computed based on  our knowledge of the shape of $A$. First, we generate a set of $T$ points, denoted by $\V{y}_1,\ldots,\V{y}_{T}$ sampled uniformly over $A$. We assume that the first $N$ points are identical to the ones of the previous graph, such that $\V{y}_1= \V{x}_1,\ldots,\V{y}_N= \V{x}_N$. 
Each node $v_i^A$ corresponds to one random location $\V{y}_i$. The weight between two nodes $v_i^A,v_j^A \in V^{A}$ is computed by a normalized Gaussian kernel with $\V{y}_i,\V{y}_j$ as input.

For both graphs, we compute a symmetric normalized Laplacian matrix, given by
\begin{eqnarray}\label{eq:rw_laplacian}
\V{L}^{S} = \left(\V{D}^{S} \right)^{-1/2} \left( \V{D}^{S} -\V{W}^{S} \right)\left(\V{D}^{S}\right)^{-1/2} \\
\nonumber
\V{L}^{A} = \left(\V{D}^{A}\right)^{-1/2} \left( \V{D}^A- \V{W}^{A}\right) \left(\V{D}^{A}\right)^{-1/2},
\end{eqnarray}
where $\V{D}^{S}$ and $\V{D}^{A}$ are diagonal matrices with  elements $\V{D}^{S}_{i,i}=\sum_j \V{W}^{S}_{i,j}$ and $\V{D}^{A}_{i,i}=\sum_j \V{W}^{A}_{i,j}$, respectively. Following \cite{belkin2003laplacian,coifman2006diffusion}. Let $\V{u}_i^{S},\V{u}_i^{A}$ be, respectively, the eigenvectors of $\V{L}^{A}$ and   $\V{L}^{S}$ that correspond to the $i$-th smallest eigenvalues.
We compute a new representation for the nodes in both graphs using the eigenvectors of the respective Laplacian matrix by
\begin{eqnarray}
\label{eq:embedding}
\V{\phi}_S(\V{S}^i) = [ (\V{u}^{S}_1)_i,\ldots,  (\V{u}^{S}_d)_i]\in \mathbb{R}^d \\ \nonumber
\V{\phi}_A(\V{y}_i) = [ (\V{u}^{A}_1)_i,\ldots,  (\V{u}^{A}_l)_i] \in \mathbb{R}^l,
\end{eqnarray}
Our approach builds upon the relation between the two graphs $\GS$ and $\GA$ and their representation based on their spectrum. As shown in \cite{hein2007graph,singer2006graph,trillos2020error,belkin2003laplacian,dunson2021spectral}, 
given a sufficiently large number of samples and signals from the area $A$, the eigenvectors $\{\V{u}_i^{A}\}$ and $\{\V{u}_i^{S}\}$ converge to the eigenfunctions of the Laplace-Beltrami operator on their respective manifolds. These eigenfunctions capture important geometric information of the underlying manifold, see \cite{belkin2003laplacian,coifman2006diffusion}. 

\cref{assum:smoothness} implies that the signals $\V{s}^i_j$ are fully determined by their location $\V{x_i}$. However, the eigenvectors of the signal graph $\GS$ can be very different from the corresponding eigenvectors of the area graph $\GA$. Let us give a concrete example. Assume that $\A\equiv[0,1]^2$. For the graphs $\GA$ and  $\GS$ we sampled a set of $M=1000$ and $T=1000$ points uniformly at random over the area $A$. The high dimensional signals are given by
\begin{equation}\label{eq:motivating_example}
\V{S}^i = \frac{1}{\|\V{x}_i-\V{r}_0\|^2}\V{B}\V{x}_i,  
\end{equation}
where $\V{B}\in \mathbb{R}^{p\times 2}$ is a random matrix and $\V{r}_0\in \mathbb{R}^2$ is some point outside $A$. The term $1/\|\V{x}_i-\V{r}_0\|$ can be understood as the result of a power decay due to the distance to a receiver. This factor implies that points with a similar distance to $\V{r}_0$ will be more strongly connected than points at different distances, which distorts the original manifold structure.

 \begin{figure}[t]
    \centering
    \includegraphics[width=.7\columnwidth]{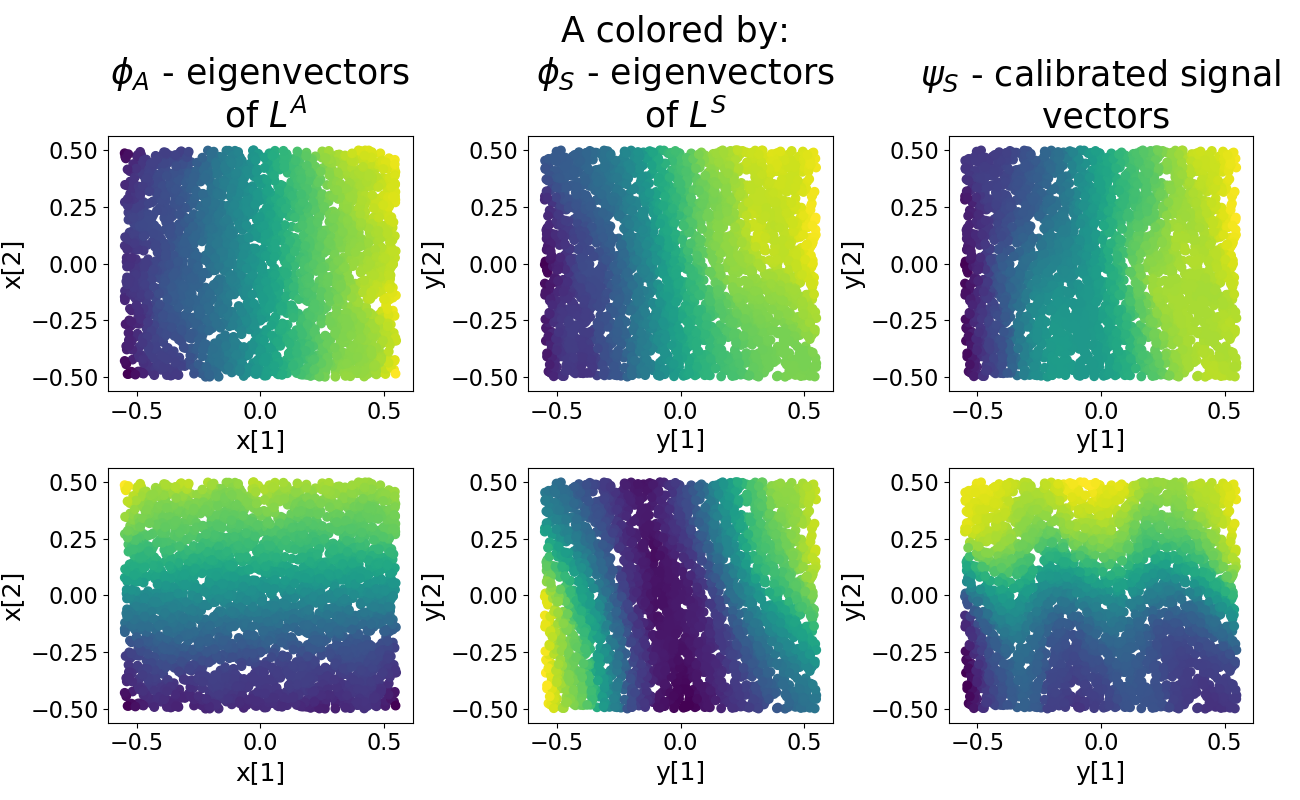}
    \caption{Signals simulated according to Eq. \eqref{eq:motivating_example} in  \cref{sec:solution}. Left: Points in $\A$ colored by the two eigenvectors that correspond to the smallest eigenvalues of $\V{L}^A$. Middle: $\A$ colored by the two eigenvectors of $\V{L}^S$, Right: The calibrated eigenvectors of the signals using Eq. \eqref{eq:calibrated_embedding_form}}
    \label{fig:motivation1}
\end{figure}

The left column of \cref{fig:motivation1} shows the two leading eigenvectors of the area graph $\GA$. Since $A$ is a square, each of the two leading eigenvectors is a cosine function of one of its coordinates. The middle column shows the two leading eigenvectors of $\GS$. It is evident that these vectors depend mainly on the radial effect caused by the non linear function $1/\|\V{x}_i-\V{r}_0\|^2$. Thus, the challenge is to calibrate these eigenvectors and extract the information relevant for localization from the signal's graph. The process computing a calibration matrix is described in the following section. 

\paragraph*{Matching spectral representation.}
Our goal is to extract from the signal embedding $\V{\phi}_S$ a \textit{calibrated representation} that is consistent with the area $A$, and thus useful for localization. We denote the calibrated representation function by $\V{\psi}_S$. 
On the one hand, $\V{\psi}_S$ should 
preserve the geometry of the signals, and thus be smooth with respect to the signal graph $G^S$.
On the other hand, for the points $i = 1,\ldots N$ whose location $\V{x}_i = \V{y}_i$ is known, we would like the calibrated representation  $\V{\psi}_S (\V{S}^i)$ to be similar to the corresponding representation of the locations $\V{\phi}_A(\V{y}_i)$.  This motivates the following optimization problem: 
\begin{equation}\nonumber
\V{\psi}_S = \argmin_{\V{\psi}= [\V{\psi}_1,\ldots,\V{\psi}_M]^{\top}\in \mathbb{R}^{M\times l}}
\frac{1}{N}\sum_{i=1}^N \left\|  \V{\phi}_A(\V{y}_i)- \V{\psi}_i  \right\|^2 
 + \frac{\lambda}{d} \cdot  \text{Tr}\left\{ \V{\psi}^{\top} \V{L}^{S} \V{\psi} \right\}.  \label{eq:compute_c}
\end{equation}

\begin{figure}[ht]
    \centering
    \includegraphics[width=.3\columnwidth]{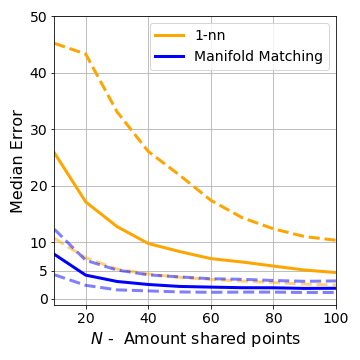}
    \includegraphics[width=.3\columnwidth]{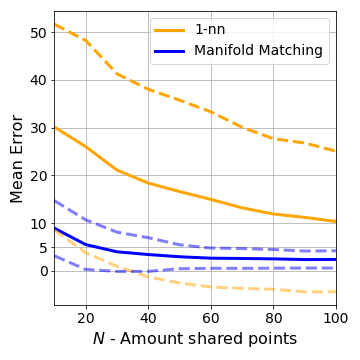}
    \includegraphics[width=.3\columnwidth]{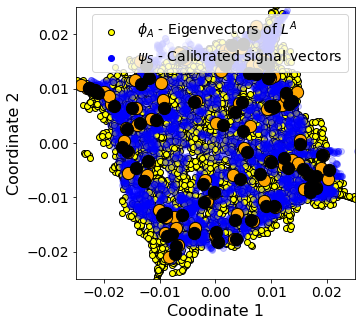}

    \caption{Experiment described in \cref{sec:experiment1}. Left: The median error of our Manifold Matching algorithm and the Naive $1-$nearest neighbor algorithm. In dashed line we supply the $25\%$ and $75\%$ percentile error. Middle: The mean error with both algorithms. In dashed line we supply the standard deviation around the mean. Right: The two leading eigenvectors of $L^A$ along the first two coordinates of the calibrated signals vector. In orange and black we supply the calibrated points embeddings of $\phi_A$ and $\psi_S$, respectively. }
    \label{fig:experiment1}
\end{figure}

The right term in Eq. \eqref{eq:compute_c} is a regularization term, which gives a penalty for using high-frequency eigenvectors from the spectrum of $\V{L}^S$. Similar regularization terms appear in several problems that include graph based regularizations, see for example \cite{belkin2004regularization,fergus2009semi,li2020graph}. 
In a typical setting, the solution of Eq. \eqref{eq:compute_c} is a linear function of a maximum of $N$ eigenvectors with the smallest eigenvalues of $L_S$. To improve robustness, we limit the new representation $\V{\psi}_S$ to be a linear combination of the $d<N$ eigenvectors with the smallest eigenvalues,
\begin{equation}\label{eq:calibration_matrix}
\V{\psi}_S (\V{S}^j) = \V{C} \cdot \V{\phi}_S(\V{S}^j), 
\end{equation}
where $\V{C} \in \R^{l \times d}$ is the calibration matrix. 
We slightly abuse notation by denoting the
signals embedding matrix by $\V{\phi}_S\equiv [\V{\phi}_S(\V{S}^1),\ldots, \V{\phi}_S(\V{S}^{M})]^{\top} \in\mathbb{R}^{ M \times d}$.
Inserting Eq. \eqref{eq:calibration_matrix} to Eq. \eqref{eq:compute_c} yields a quadratic optimization problem with the following closed form solution,
\begin{equation} \nonumber
 \V{C} = \sum_{i=1}^N \V{\phi}_A (\V{y}_i)\V{\phi}_S (\V{S}^j)^T   
 \cdot \left( \sum_{j=1}^N \V{\phi}_S (\V{S}^j)\V{\phi}_S (\V{S}^j)^T   +\frac{\lambda \cdot N}{d} \V{\phi}_S^{\top} \V{L}^S \V{\phi}_S \right)^{-1} .
 \label{eq:calibrated_embedding_form}
\end{equation}
The right column in  \cref{fig:motivation1} shows the first two calibrated signal vectors $\V{\psi}_S$ for the artificial signal given in Eq. \eqref{eq:motivating_example}. The number of shared points is $N=10$, and the dimensions of the signals' and locations' representation is equal to $d=8$ and $l = 2$,  respectively.
Though the calibration is done by a relatively small number of points, the new vectors indeed extract the information on the coordinates $\V{x}_i$ while leaving out other dependencies, such as the radial term. Our final step is to use the calibrated vectors in order to obtain an accurate localization for all signals. 

\begin{remark}\label{re:independent_harmonics}
For the vectors $\V{\phi}_A$, we select the $d$ eigenvectors with the smallest eigenvalue of the Laplacian matrix. These are the \textit{smoothest} orthogonal vectors with respect to the area graph. A possible alternative for $\V{\phi}_A$ is to compute the leading independent harmonics of $A$. These consist of a subset of eigenvectors that are \textit{non-redundant}, in the sense that each captures a different coordinate in the original manifold, see \cite{blau2017non,dsilva2018parsimonious}.
A simple example where such an approach may be beneficial is a long rectangular area, where the eigenvectors with the smallest eigenvalues will all encode the same coordinate. In such a case, taking independent harmonics may guarantee that we obtain non-redundant eigenvectors, that are sufficient for localization. 
\end{remark}

\paragraph*{Localization via 1-nearest neighbor}

Finally, based on the calibrated coordinates, we estimate the location of any unknown point $x_i$ by a simple $1$-nn search,  
\begin{eqnarray}
 \hat {\V{x}}_i = \argmin_{\V{y}_j} \| \V{\phi}_A(\V{y}_j)-  \V{\psi}_S (\V{S}^i)\|.
 \label{eq:estimate_loc}
 \end{eqnarray}
In other words, we search for the point $\V{y}_j$ that after calibration, has the closest representation to $\V{S}^i$.
Algorithm \ref{alg:localization} summarizes the different steps of our localization pipeline.

\section{Experimental results}
\label{sec:experiments}
We validate our localization approach on simulated and real datasets captured in  an office environment. In the first experiment, we use the simulated dataset generated in \cite{kupershtein2012single} by a 3D wave propagation software. The floor plan of this simulation is shown in  \cref{fig:indoor_positioning_setting}. In our second experiment, we test our approach  on a real Wifi signal dataset generated in \cite{torres2014ujiindoorloc}. The code for reproducing our results is provided as supplementary material.  


\subsection{Simulated Wifi experiment}
\label{sec:experiment1}

In our first experiment, the area $A$ is a rectangular office environment of size $80m \times  80m$ illustrated in \cref{fig:indoor_positioning_setting}. The dataset consists of $200000$ signals of dimension $p=48$ along with their transmitted locations. 
The subsets of signals $\V{S}^1,\ldots,\V{S}^M$ and the local metric function are defined similarly to  \cite{kupershtein2012single, jaffe2014single,moscovich2017minimax}. 
Specifically, we subsampled $M=8000$ random locations from the dataset ($\V{x}_1,\ldots,\V{x}_M$). For each subsampled point $\V{x}_i$, we defined a signal set $\V{S}^i\in \mathbb{C}^{K\times p}$ as the concatenation of the $K=80$ signals with the closest transmitted location to $\V{x}_i$, within $1m$ of the transmitted location. 
For the construction of the signal graph $\GS$ we follow the similarity criteria between subsets of signals derived in  \cite{kupershtein2012single}, see more details in \cref{sec:tech_details}.

To construct $\GA$, we sampled $T=8000$ points over $A$. The embedding dimensions are set to $l=10$ for the area graph and $d=30$ for the signal graph.
The right panel in \cref{fig:experiment1} shows the two leading eigenfunctions of $L^A$ in yellow, and the calibrated eigenfunctions in blue. The figure shows that the calibration process matches the signal eigenfuncions to the area eigenfunctions.   

The left two panels in \cref{fig:experiment1} show the localization error of our algorithm as a function of the number of shared points. As baseline, we compared the 
performance to a $1$nn classifier that assigns each signal to the closest labeled point. Our semi-supervised approach shows a clear advantage. For $N = 80$ the median error of our estimate is equal to $1.97m$m and the mean error is $2.52$. For additional comparison, in \cite{moscovich2017minimax} the same dataset was tested, and a mean error of $2.5m$ was achieved with $~700$ labeled points. We thus reach a similar accuracy with around $11\%$ of the labeled points.



\subsection{Wifi received signal strength experiment}
\label{sec:experiment2}

\begin{figure}[ht!]
    \centering
    \includegraphics[width = 0.45\linewidth]{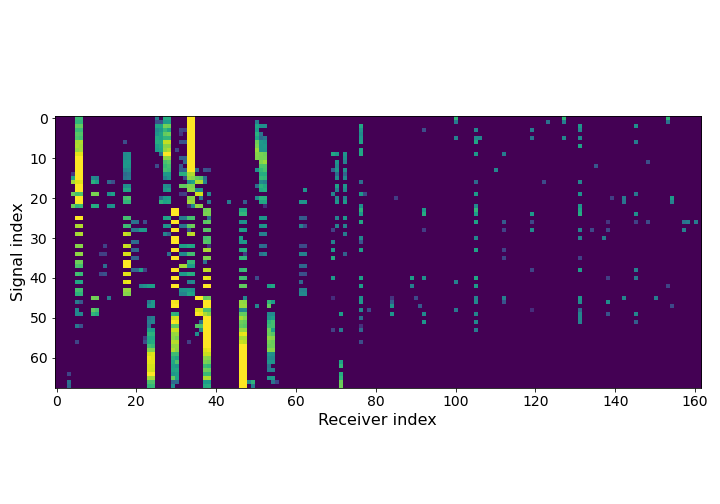}
    \includegraphics[width = .45\linewidth ]{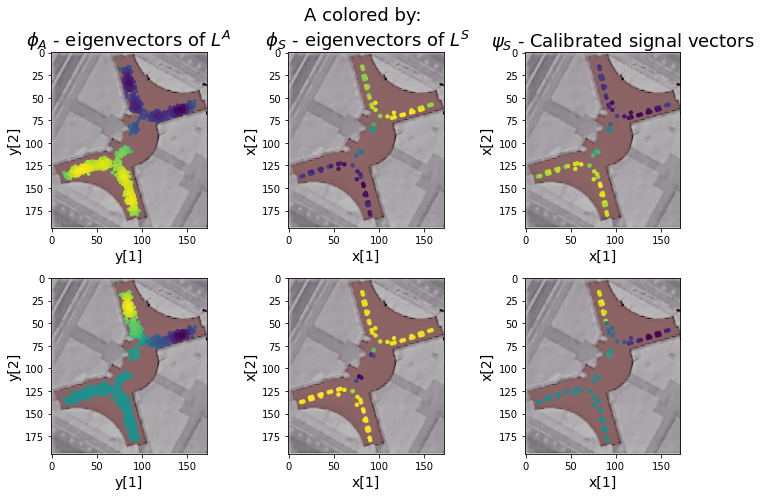}
    \caption{Experiment described in \cref{sec:experiment2}. Left: A matrix containing an average signal strength from each point to the set of receivers. Right: The  
    background is one of the buildings of  Universitat Jaume (UJ). The left column shows simulated points in the are $A$, colored by the first two coordinates of the spectral embedding of $\GA$, $\phi_A$. The middle column shows locations of transmitted signals, colored by the first two coordinates of the spectral embedding of $\GS$, $\phi_S$. In the right column the same locations are now colored by the first two coordinates of the calibrated signal representation, $\psi_S$.}
    \label{fig:experiment2}
\end{figure}

The second experiment is based on a Wifi signal dataset generated in \cite{torres2014ujiindoorloc}. The dataset consists of the received signal strength measured by $p = 520$ receivers, in an area of $12$ floors of $3$ buildings ($4$ floors each). At each point $\V{x}_i$, $K$ messages were transmitted and captured. Let  $\V{S}^i = [\V{s}_1^i,\ldots,\V{s}_K^i]^\top \in \{0,1\}^{K \times 520 }$ 
denote a matrix whose elements contain the signal strength of $K$ messages transmitted from  locations  $\V{x}_i$. 
For the signal graph $\GS$ we computed the median signal $\bar{ \V{s}}^i \in \mathbb R^{520}$, equal to
\[
\bar{ \V{s}}^i = \text{median}( 
 \V{s}_1^i, \ldots, \V{s}_K^i ).
\]
The left panel in Figure \ref{fig:experiment2} shows the values of $\bar{ \V{s}} _j$ for different locations $\V{x}_j$. For this floor, only $160$ out of $520$ receivers captured one of the message.

\begin{figure}[ht!]
    \centering
    \includegraphics[width=.3\columnwidth]{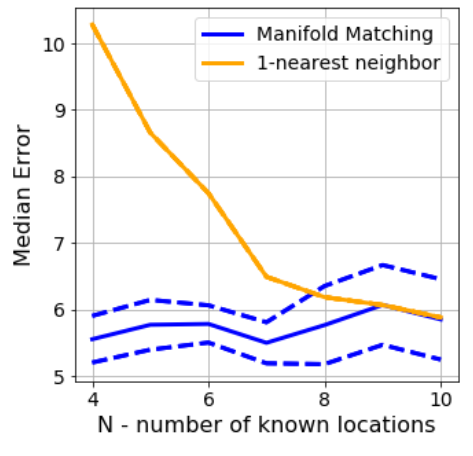}
    \includegraphics[width=.3\columnwidth]{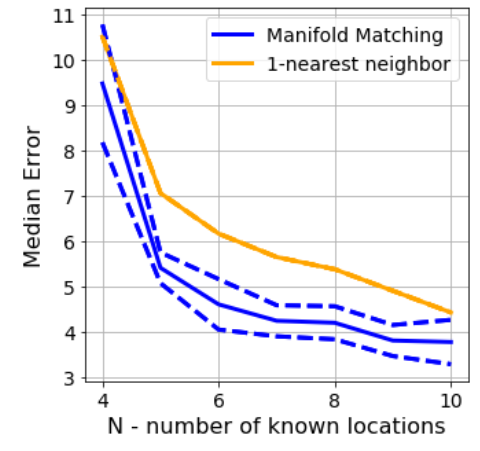}
    \includegraphics[width=.3 \columnwidth]{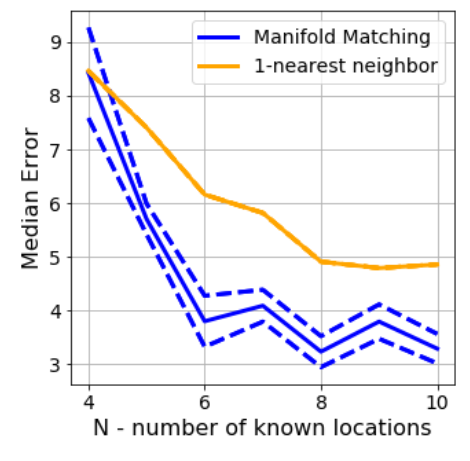}
    \caption{Median error of the manifold matching approach as a function of the number of shared points. The dashed lines mark the standard deviation error of $10$ independent runs with different choices of shared points. The environments are illustrated in Figures \ref{fig:indoor_positioning_setting} and \ref{fig:experiment2_otherbuilding}. Left panel: building index $2$ floor index $3$, (see illustration in Figure \ref{fig:experiment2_otherbuilding}.) The parameters were set to $\lambda=0.01$, $l=4$, and $d=6$.  Middle and right panels: building index 0, floor indices 3 and 2 (see \cref{fig:experiment2}). Parameters set to  $l=6$, $d=10$ and $\lambda = 0.01$. }
    \label{fig:experiment2_results}
\end{figure}

For the graph $\GS$ the weight matrix is computed by a normalized Gaussian kernel $K(\bar{ \V{s}}_i,\bar{ \V{s}}_j)$. For the graph $\GA$, 
we sampled $1000$ points in the area of the transmitted signals. The middle panel of  \cref{fig:experiment2} shows, as background, one of the floors from which signals were sent. The left column shows the $1000$ simulated points, colored by their values in two eigenvectors of $\GA$. Similarly, the middle column shows the locations of the actual signals, colored by their values in two eigenvectors of $\GS$. The right column shows the calibrated vectors with $N=10$ calibration points. The location of the $N$ shared points was chosen by running k-means on the signal dataset. 
Figure \ref{fig:experiment2_results} shows the median error as a function of the number of shared points for three different floors.   Similarly to the first experiment, we compare our performance to a $1$nn classifier. The results show a clear advantage and demonstrate the importance of learning the intrinsic structure of the signals for localization. 


\section{Discussion}
\begin{figure}[htb]
    \centering
    \includegraphics[width=.3\columnwidth,trim=0mm 0 0 0mm,clip]{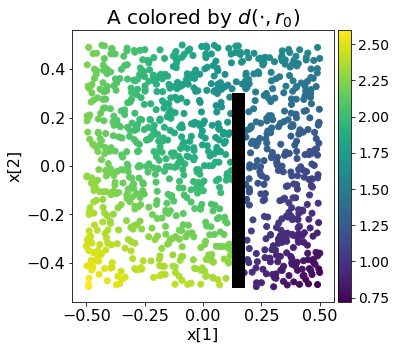}
    \includegraphics[width=.24\columnwidth]{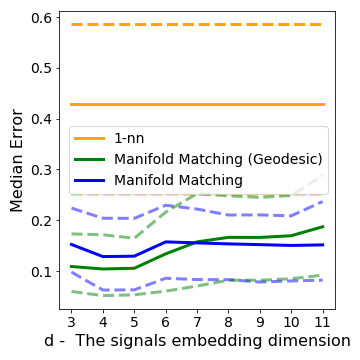}
    \includegraphics[width=.44\columnwidth]{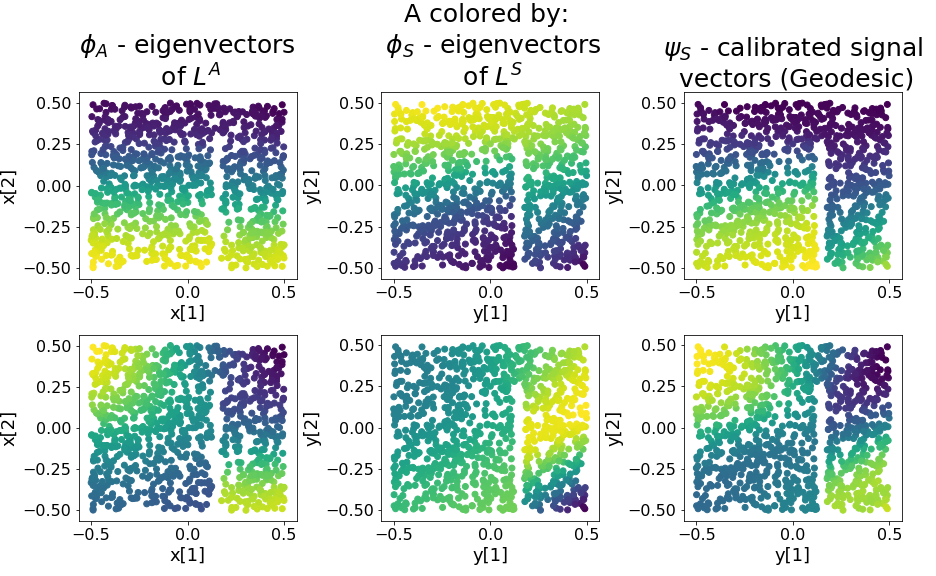}
    \caption{Left panel: random locations in an indoor environment colored by their geodesic distance to a point $\V{r}_0$. The black line illustrates a wall. 
    Middle panel: comparison of the accuracy of indoor positioning as a function of the embedding dimension $d$. The dashed lines are the 25th and 75th percentile. Right panel: - 
    Coloring the locations by the leading two coordinates of the spectral embedding. Left column - area embedding, middle column - signal embedding, and right column - calibrated signal  embedding. 
    }
    \label{fig:geodesic}
\end{figure}

\label{sec:discussion}

\paragraph{Choice of parameters.}
Algorithm \ref{alg:localization} accepts, as input, three parameters whose choice may impact its accuracy: the dimensions  $l$,$d$ of the signal and area embeddings, and the regularization parameters $\lambda$. 
One simple approach for selecting these parameters is using the $N$ shared points for cross validation by computing the calibrated vectors and testing the performance on non-overlapping subsets of the $N$ shared points.

\begin{figure}[b]
    \centering
    \includegraphics[width = 0.3\linewidth]{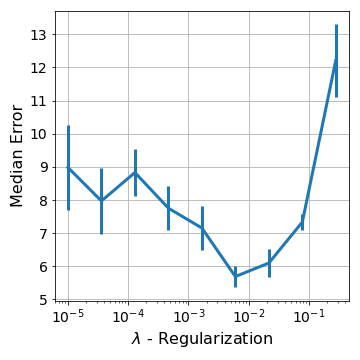}
    \includegraphics[width = 0.3\linewidth]{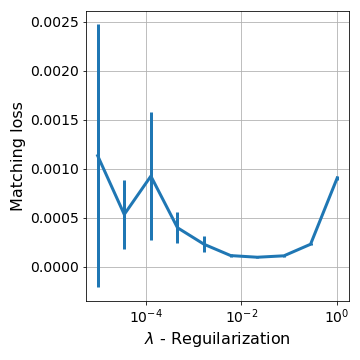} 
    \caption{Choice of $\lambda$ for the data used in our Wifi experiment from \cite{torres2014ujiindoorloc}.  Left: median localization error as a function of $\lambda$. Right: Matching loss function defined in \eqref{eq:icp_loss} as a function of $\lambda$.}
    \label{fig:lambda_icp_choice}
\end{figure}

Applying cross-validation may be problematic, however, when the number of points $N$ is small. Inspired by the Iterative Closest Point algorithm \cite{besl1992method}, 
we propose an alternative approach that measures how well the area manifold and the signal manifold match after calibration, see for example Figure \ref{fig:experiment1}. For determining $\lambda$ we define the following loss function:
\begin{align}\label{eq:icp_loss}
l(\lambda) = \frac{1}{T}\sum_{i=1}^T \min_j \| \V{\phi}_A( \V{y}_i)- \V{\psi}_S(\V{S}^j)\|^2 
+ \frac{1}{M} \sum_{j=1}^M \min_i \| \V{\phi}_A(\V{y}_i)- \V{\psi}_S(\V{S}^j)\|^2.      
\end{align}
We propose to determine $\lambda$ by minimizing the loss in Eq. \eqref{eq:icp_loss}. 
We tested this approach for one of the datasets in \cite{torres2014ujiindoorloc}, which was used for our Wifi experiment. 
The left panel in Figure \ref{fig:lambda_icp_choice} (Appendix) shows the performance of our Manifold Matching approach as a function of $\lambda$. The right panel shows the value of $l(\lambda)$ as defined in Eq. \eqref{eq:icp_loss}. The two curves are highly correlated, and achieve a minima around $\lambda \approx 0.01$. This suggests that the matching based loss function is a valid way to choose $\lambda$ for cases where the number of shared points is small. This approach may be used to select the embedding dimension $d$ as well. The choice of $l$ by this approach is more problematic as the loss increases with the dimension $l$.

\paragraph{Integrating additional signals.} 
Once a match has been found between the signal space and the location space, the location of a new signal $\V{S}$ can be obtained by finding the signal $\V{S}^j$ with the most similar features. Let $\hat{\V{x}}(S)$ be the estimated location of $\V{S}$, then 
\begin{eqnarray*}
\hat{\V{x}}(\V{S}) &=& \hat{\V{x}}_k \\
k &=& \underset{j\in M}{\text{argmax}} \,\,K(\V{S},\V{S}^j).
\end{eqnarray*}
Other techniques can be employed for the out-of-sample-extension, an example for such is the Nystorm extension to infer the new signal embedding based on the known signals as discussed in \cite{lafon2006data}. In some cases, one would like to integrate incoming signals so as to improve the quality of the original data. To that end, recent works implemented neural network architectures to compute spectral embeddings of new data points \cite{mishne2019diffusion,shaham2018spectralnet}. 
\label{sec:analysis}

\paragraph{Discontinuity.}
The indoor multipath phenomena, in which a  signal propagates along different trajectories, poses multiple problems for indoor localization. 
One such problem is that the features of a received signal transmitted from two sides of a concrete wall may be very different. 
Disregarding such effects may harm the performance of our approach as the smoothness assumption (3.3) is strongly violated. To illustrate the effect of a physical barrier, we repeated the experiment described in section \ref{sec:solution},  but with a slight modification. We assume that there is a vertical wall that causes a \textit{diffraction} pattern, see Figure \ref{fig:geodesic}. To model diffraction, we replace the Euclidean distance $\|x-r_0\|$ in Eq. \eqref{eq:motivating_example} with a geodesic distance, denoted by $d(\V{x},r_0)$. Specifically, the signal space is computed based on the following transformation,
\begin{eqnarray}
\V{S}^i=  \frac{1}{d(\V{x}_i,\V{r}_0)^2}\V{B}\V{x}_i.
\end{eqnarray}
Assuming the location of the barrier is known, it can be taken into account when computing the area graph $G^A$. This can be done by using the geodesic distance for the area graph $G^A$. This will effectively remove connections between points at two sides of a wall. 
The right panels of \cref{fig:geodesic} show the spectral embedding of of $A$ (left column), the signal embedding (middle column) and the calibrated signal (right column). Taking into account the barrier simplifies the calibration process as the discontinuity appears both in the eigenvecors of $L^A$ and those of $L^S$. 

\paragraph{Independent harmonics.} In remark \ref{re:independent_harmonics}, we mentioned the possibility of using the leading independent harmonics instead of the eigenvectors with the smallest eigenvalue. 
This may insure that we gain sufficient information for each coordinate of the area $A$. A related question for future research is the number of eigenvectors or independent harmonics one requires to obtain accurate localization. On the one hand, a single eigenvector for each coordinate may simplify calibration and thus the required number of shared points. On the other hand, a localization process based on a single eigenvector for each coordinate may be sensitive to noise. For example, the first harmonic of a rectangular area is a cosine function with low frequency, which has a very low gradient in areas close to the borders of $A$. Thus, even a small perturbation can cause a large localization error. Taking additional harmonics where the gradient is higher may thus is expected to increase the accuracy in these areas. 

\section{Acknowledgment}
EP and IGK were partially supported by DARPA (ATLAS program, Dr. J. Zhou) and the US AFPSR (Dr. Fariba Fahroo). EP was partially supported by the Center for Interdisciplinary Data Science Research at the Hebrew University (CIDR). We would like to thank Mati Wax, 
Boris Landa and Ronald Coifman for for useful and insightful discussions.


\printbibliography

@inproceedings{moscovich2017minimax,
  title={Minimax-optimal semi-supervised regression on unknown manifolds},
  author={Moscovich, Amit and Jaffe, Ariel and Boaz, Nadler},
  booktitle={Artificial Intelligence and Statistics},
  pages={933--942},
  year={2017},
  organization={PMLR}
}

@article{li2020graph,
  title={Graph-Based Regularization for Regression Problems with Alignment and Highly Correlated Designs},
  author={Li, Yuan and Mark, Benjamin and Raskutti, Garvesh and Willett, Rebecca and Song, Hyebin and Neiman, David},
  journal={SIAM journal on mathematics of data science},
  volume={2},
  number={2},
  pages={480--504},
  year={2020},
  publisher={SIAM}
}

@article{trillos2020error,
  title={Error estimates for spectral convergence of the graph Laplacian on random geometric graphs toward the Laplace--Beltrami operator},
  author={Trillos, Nicol{`a}s Garc{`}a and Gerlach, Moritz and Hein, Matthias and Slepcev, Dejan},
  journal={Foundations of Computational Mathematics},
  volume={20},
  number={4},
  pages={827--887},
  year={2020},
  publisher={Springer}
}

@article{hein2007graph,
  title={Graph laplacians and their convergence on random neighborhood graphs.},
  author={Hein, Matthias and Audibert, Jean-Yves and Luxburg, Ulrike von},
  journal={Journal of Machine Learning Research},
  volume={8},
  number={6},
  year={2007}
}

@article{kupershtein2012single,
  title={Single-site emitter localization via multipath fingerprinting},
  author={Kupershtein, Evgeny and Wax, Mati and Cohen, Israel},
  journal={IEEE Transactions on signal processing},
  volume={61},
  number={1},
  pages={10--21},
  year={2012},
  publisher={IEEE}
}

@article{jaffe2014single,
  title={Single-site localization via maximum discrimination multipath fingerprinting},
  author={Jaffe, Ariel and Wax, Mati},
  journal={IEEE Transactions on Signal Processing},
  volume={62},
  number={7},
  pages={1718--1728},
  year={2014},
  publisher={IEEE}
}

@article{zafari2019survey,
  title={A survey of indoor localization systems and technologies},
  author={Zafari, Faheem and Gkelias, Athanasios and Leung, Kin K},
  journal={IEEE Communications Surveys \& Tutorials},
  volume={21},
  number={3},
  pages={2568--2599},
  year={2019},
  publisher={IEEE}
}

@inproceedings{ni2003landmarc,
  title={LANDMARC: Indoor location sensing using active RFID},
  author={Ni, Lionel M and Liu, Yunhao and Lau, Yiu Cho and Patil, Abhishek P},
  booktitle={Proceedings of the First IEEE International Conference on Pervasive Computing and Communications, 2003.(PerCom 2003).},
  pages={407--415},
  year={2003},
  organization={IEEE}
}

@article{khalajmehrabadi2017modern,
  title={Modern WLAN fingerprinting indoor positioning methods and deployment challenges},
  author={Khalajmehrabadi, Ali and Gatsis, Nikolaos and Akopian, David},
  journal={IEEE Communications Surveys \& Tutorials},
  volume={19},
  number={3},
  pages={1974--2002},
  year={2017},
  publisher={IEEE}
}

@article{belkin2003laplacian,
  title={Laplacian eigenmaps for dimensionality reduction and data representation},
  author={Belkin, Mikhail and Niyogi, Partha},
  journal={Neural computation},
  volume={15},
  number={6},
  pages={1373--1396},
  year={2003},
  publisher={MIT Press}
}

@article{coifman2006diffusion,
  title={Diffusion maps},
  author={Coifman, Ronald R and Lafon, St{\'e}phane},
  journal={Applied and computational harmonic analysis},
  volume={21},
  number={1},
  pages={5--30},
  year={2006},
  publisher={Elsevier}
}

@inproceedings{witrisal2012performance,
  title={Performance bounds for multipath-assisted indoor navigation and tracking (MINT)},
  author={Witrisal, Klaus and Meissner, Paul},
  booktitle={2012 IEEE International Conference on Communications (ICC)},
  pages={4321--4325},
  year={2012},
  organization={IEEE}
}

@article{liu2014survey,
  title={Survey of wireless based indoor localization technologies},
  author={Liu, Junjie and Jain, R},
  journal={Department of Science \& Engineering, Washington University},
  year={2014}
}

@article{yang2013rssi,
  title={From RSSI to CSI: Indoor localization via channel response},
  author={Yang, Zheng and Zhou, Zimu and Liu, Yunhao},
  journal={ACM Computing Surveys (CSUR)},
  volume={46},
  number={2},
  pages={1--32},
  year={2013},
  publisher={ACM New York, NY, USA}
}

@article{wang2016tdoa,
  title={TDOA positioning irrespective of source range},
  author={Wang, Yue and Ho, KC},
  journal={IEEE Transactions on Signal Processing},
  volume={65},
  number={6},
  pages={1447--1460},
  year={2016},
  publisher={IEEE}
}

@inproceedings{barsocchi2009novel,
  title={A novel approach to indoor RSSI localization by automatic calibration of the wireless propagation model},
  author={Barsocchi, Paolo and Lenzi, Stefano and Chessa, Stefano and Giunta, Gaetano},
  booktitle={VTC Spring 2009-IEEE 69th Vehicular Technology Conference},
  pages={1--5},
  year={2009},
  organization={IEEE}
}

@article{chen2017confi,
  title={ConFi: Convolutional neural networks based indoor Wi-Fi localization using channel state information},
  author={Chen, Hao and Zhang, Yifan and Li, Wei and Tao, Xiaofeng and Zhang, Ping},
  journal={IEEE Access},
  volume={5},
  pages={18066--18074},
  year={2017},
  publisher={IEEE}
}

@article{koelle2018manifold,
  title={Manifold coordinates with physical meaning},
  author={Koelle, Samson and Zhang, Hanyu and Meila, Marina and Chen, Yu-Chia},
  journal={arXiv e-prints},
  pages={arXiv--1811},
  year={2018}
}

@article{ghazvinian2021modality,
  title={Modality-Agnostic Topology Aware Localization},
  author={Ghazvinian Zanjani, Farhad and Karmanov, Ilia and Ackermann, Hanno and Dijkman, Daniel and Merlin, Simone and Welling, Max and Porikli, Fatih},
  journal={Advances in Neural Information Processing Systems},
  volume={34},
  year={2021}
}

@article{singer2006graph,
  title={From graph to manifold Laplacian: The convergence rate},
  author={Singer, Amit},
  journal={Applied and Computational Harmonic Analysis},
  volume={21},
  number={1},
  pages={128--134},
  year={2006},
  publisher={Elsevier}
}

@inproceedings{torres2014ujiindoorloc,
  title={UJIIndoorLoc: A new multi-building and multi-floor database for WLAN fingerprint-based indoor localization problems},
  author={Torres-Sospedra, Joaqu{'i}n and Montoliu, Ra{'u}l and Mart{'i}nez-Us{'o}, Adolfo and Avariento, Joan P and Arnau, Tom{'a}s J and Benedito-Bordonau, Mauri and Huerta, Joaqu{'\i}n},
  booktitle={2014 international conference on indoor positioning and indoor navigation (IPIN)},
  pages={261--270},
  year={2014},
  organization={IEEE}
}

@article{belkin2004semi,
  title={Semi-supervised learning on Riemannian manifolds},
  author={Belkin, Mikhail and Niyogi, Partha},
  journal={Machine learning},
  volume={56},
  number={1},
  pages={209--239},
  year={2004},
  publisher={Springer}
}

@article{villoutreix2017synthesizing,
  title={Synthesizing developmental trajectories},
  author={Villoutreix, Paul and And{'e}n, Joakim and Lim, Bomyi and Lu, Hang and Kevrekidis, Ioannis G and Singer, Amit and Shvartsman, Stanislav Y},
  journal={PLoS computational biology},
  volume={13},
  number={9},
  pages={e1005742},
  year={2017},
  publisher={Public Library of Science San Francisco, CA USA}
}

@inproceedings{fergus2009semi,
  title={Semi-Supervised Learning in Gigantic Image Collections.},
  author={Fergus, Rob and Weiss, Yair and Torralba, Antonio},
  booktitle={NIPS},
  volume={1},
  pages={2},
  year={2009},
  organization={Citeseer}
}

@article{peterfreund2020local,
  title={Local conformal autoencoder for standardized data coordinates},
  author={Peterfreund, Erez and Lindenbaum, Ofir and Dietrich, Felix and Bertalan, Tom and Gavish, Matan and Kevrekidis, Ioannis G and Coifman, Ronald R},
  journal={Proceedings of the National Academy of Sciences},
  volume={117},
  number={49},
  pages={30918--30927},
  year={2020},
  publisher={National Acad Sciences}
}

@inproceedings{zhou2004learning,
  title={Learning with local and global consistency},
  author={Zhou, Dengyong and Bousquet, Olivier and Lal, Thomas N and Weston, Jason and Scholkopf, Bernhard},
  booktitle={Advances in neural information processing systems},
  pages={321--328},
  year={2004}
}

@incollection{zhou2014semi,
  title={Semi-supervised learning},
  author={Zhou, Xueyuan and Belkin, Mikhail},
  booktitle={Academic Press Library in Signal Processing},
  volume={1},
  pages={1239--1269},
  year={2014},
  publisher={Elsevier}
}

@article{dunson2021spectral,
  title={Spectral convergence of graph Laplacian and Heat kernel reconstruction in L infinity from random samples},
  author={Dunson, David B and Wu, Hau-Tieng and Wu, Nan},
  journal={Applied and Computational Harmonic Analysis},
  year={2021},
  publisher={Elsevier}
}

@inproceedings{
shaham2018spectralnet,
title={SpectralNet: Spectral Clustering using Deep Neural Networks},
author={Uri Shaham and Kelly Stanton and Henry Li and Ronen Basri and Boaz Nadler and Yuval Kluger},
booktitle={International Conference on Learning Representations},
year={2018},
}

@article{mishne2019diffusion,
  title={Diffusion nets},
  author={Mishne, Gal and Shaham, Uri and Cloninger, Alexander and Cohen, Israel},
  journal={Applied and Computational Harmonic Analysis},
  volume={47},
  number={2},
  pages={259--285},
  year={2019},
  publisher={Elsevier}
}

@article{singer2008non,
  title={Non-linear independent component analysis with diffusion maps},
  author={Singer, Amit and Coifman, Ronald R},
  journal={Applied and Computational Harmonic Analysis},
  volume={25},
  number={2},
  pages={226--239},
  year={2008},
  publisher={Elsevier}
}

@inproceedings{litany2017fully,
  title={Fully spectral partial shape matching},
  author={Litany, Or and Rodol{`a}, Emanuele and Bronstein, Alexander M and Bronstein, Michael M},
  booktitle={Computer Graphics Forum},
  volume={36},
  number={2},
  pages={247--258},
  year={2017},
  organization={Wiley Online Library}
}

@inproceedings{rodola2017partial,
  title={Partial functional correspondence},
  author={Rodol{`a}, Emanuele and Cosmo, Luca and Bronstein, Michael M and Torsello, Andrea and Cremers, Daniel},
  booktitle={Computer Graphics Forum},
  volume={36},
  number={1},
  pages={222--236},
  year={2017},
  organization={Wiley Online Library}
}

@article{dsilva2018parsimonious,
  title={Parsimonious representation of nonlinear dynamical systems through manifold learning: A chemotaxis case study},
  author={Dsilva, Carmeline J and Talmon, Ronen and Coifman, Ronald R and Kevrekidis, Ioannis G},
  journal={Applied and Computational Harmonic Analysis},
  volume={44},
  number={3},
  pages={759--773},
  year={2018},
  publisher={Elsevier}
}

@inproceedings{blau2017non,
  title={Non-redundant spectral dimensionality reduction},
  author={Blau, Yochai and Michaeli, Tomer},
  booktitle={Joint European Conference on Machine Learning and Knowledge Discovery in Databases},
  pages={256--271},
  year={2017},
  organization={Springer}
}

@article{belkin2006manifold,
  title={Manifold regularization: A geometric framework for learning from labeled and unlabeled examples.},
  author={Belkin, Mikhail and Niyogi, Partha and Sindhwani, Vikas},
  journal={Journal of machine learning research},
  volume={7},
  number={11},
  year={2006}
}

@inproceedings{belkin2004regularization,
  title={Regularization and semi-supervised learning on large graphs},
  author={Belkin, Mikhail and Matveeva, Irina and Niyogi, Partha},
  booktitle={International Conference on Computational Learning Theory},
  pages={624--638},
  year={2004},
  organization={Springer}
}

@inproceedings{besl1992method,
  title={Method for registration of 3-D shapes},
  author={Besl, Paul J and McKay, Neil D},
  booktitle={Sensor fusion IV: control paradigms and data structures},
  volume={1611},
  pages={586--606},
  year={1992},
  organization={International Society for Optics and Photonics}
}

@inproceedings{NIPS2004_40173ea4,
 author = {Zelnik-manor, Lihi and Perona, Pietro},
 booktitle = {Advances in Neural Information Processing Systems},
 editor = {L. Saul and Y. Weiss and L. Bottou},
 pages = {},
 publisher = {MIT Press},
 title = {Self-Tuning Spectral Clustering},
 volume = {17},
 year = {2005}
}

@article{lafon2006data,
  title={Data fusion and multicue data matching by diffusion maps},
  author={Lafon, Stephane and Keller, Yosi and Coifman, Ronald R},
  journal={IEEE Transactions on pattern analysis and machine intelligence},
  volume={28},
  number={11},
  pages={1784--1797},
  year={2006},
  publisher={IEEE}
}

\appendix
\onecolumn
\section{Computing the Kernel matrix}
The normalized Gaussian kernel used in this paper is equal to
\begin{eqnarray*}
K(x_i,x_j) =  \frac{exp(-\|x_i-x_j\|_2^2 /\sigma^2 )}{ \sum_k exp(-\|x_i-x_k\|_2^2/\sigma^2) \cdot \sum_t exp(-\|x_j-x_t\|_2^2/\sigma^2)}.
\end{eqnarray*}
In all graphs, except the signal graph in experiment 1 and the discontinuity experiment in Section \ref{sec:discussion} the bandwidth $\sigma$ was computed by 
\begin{eqnarray}
\sigma = \max_i \|x_i -x_{k(i)}\|
\end{eqnarray}
where $k(i)$ returns the k-nearest neighbor index of $x_i$. In the experiment discussed in the discontinuity paragraph in \cref{sec:discussion}, we use a self tuning kernel bandwidth as suggest in \cite{NIPS2004_40173ea4} where
\begin{eqnarray*}
K(x_i,x_j) =  \frac{exp(-\|x_i-x_j\|_2^2 /(\sigma_i\sigma_j) )}{ \sum_k exp(-\|x_i-x_k\|_2^2/(\sigma_i\sigma_k)) \cdot \sum_t exp(-\|x_j-x_t\|_2^2/(\sigma_i\sigma_j))},\\
\sigma_i = \|x_i -x_{k(i)} \|.
\end{eqnarray*}
Details for the signal graph of experiment 1 are given in  \cref{sec:experiment1}.


\section{Computing the signal graph in experiment 1}
\label{sec:tech_details}

Following \cite{kupershtein2012single, moscovich2017minimax}, we define  for each point $\V{x}_i$ a second moment matrix $\V{R}_i$, and a corresponding projection matrix given by
\begin{eqnarray*}
\V{R}_i = \frac{1}{\|\V{S}^i \|_F^2} \V{S}^i \left(\V{S}^i\right)^{\top} \in \mathbb{C}^{p\times p}\\
\V{P}_i = \sum_{i=1}^{10} \V{v}_i \V{v}_i^T \in \mathbb{C}^{p \times p},
\end{eqnarray*}
where $\V{v}_j$ is the $j$-th leading eigenvector of $\V{R}_i$. The similarities between $\V{S}^1,\ldots,\V{S}^M$ to $\V{S^i}$ are determined by
\begin{eqnarray}\label{eq:trace_p_r}
F_i(\V{S}^j) =    \Tr(\V{R}_i^{H} \V{P}_j )  \qquad \qquad j=1,\ldots,M,
\end{eqnarray}
where $H$ is the conjugate transpose operator. 
In \cite{kupershtein2012single}, the location of an unknown point was estimated by the closest labeled point via Eq. \eqref{eq:trace_p_r}.  Here, we use Eq. \eqref{eq:trace_p_r} 
to obtain the graph $\GS$. Specifically, 
the weight matrix $\V{W}^{S}$ of $\GS$ is set to be a binary nearest neighbor matrix based on $K(\cdot,\cdot)$ defined by-  $K(\V{S}^i,\V{ S}^j)=1$ if either $\V{S}^j$ or $\V{S}^i $ is among the $10$ most similar signals according to the other signal's similarity criterion. 



\section{Building 2 in Wifi experiment}
\,
\begin{figure}[H]
    \centering
    \includegraphics[width = 0.45\linewidth]{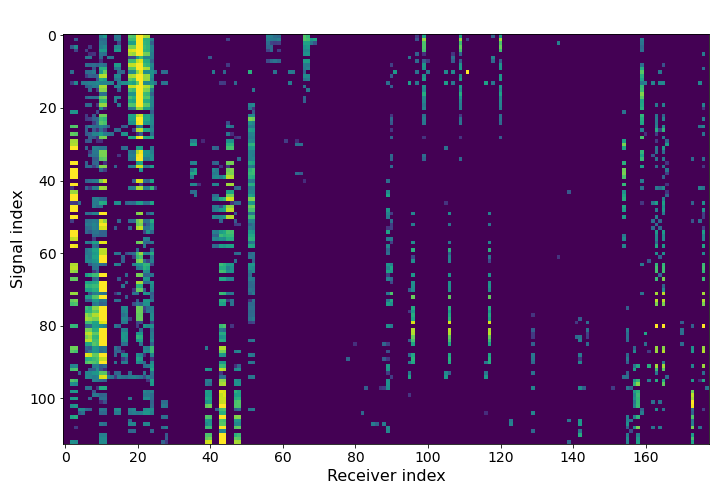}
    \includegraphics[width = 0.45\linewidth]{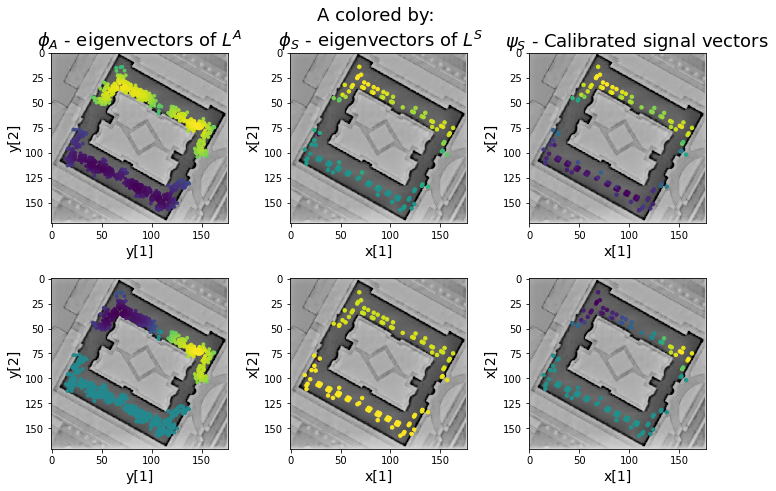}
    \caption{Signals and eigenvectors for building $2$ in \cite{torres2014ujiindoorloc}.}
    \label{fig:experiment2_otherbuilding}
\end{figure}



\end{document}